# Symmetrical bipolar electrobending deformation in acceptor-doped piezoceramics

Yi Cheng[#], Shuo Tian[#], Bin Li, Yejing Dai*

School of Materials, Sun Yat-sen University, Shenzhen, 518107, P. R. China

[#]These authors contributed equally to this work.

*Corresponding author Email: daiyj8@mail.sysu.edu.cn

**Abstract**

Since 2022, large apparent strains (>1%) with highly asymmetrical strain-electric field (*S-E*) curves have been reported in various thin piezoceramic materials, attributed to a bidirectional electric-field-induced bending (electrobending) deformation, which consistently produces convex bending along the negative electric field direction. In this study, we report a novel unidirectional electrobending behavior in acceptor-doped $K_{0.5}Na_{0.5}NbO_3$ ceramics, where convex bending always occurs along the pre-poling direction regardless of the direction of the applied electric field. This unique deformation is related to the reorientation of the $(M_{Nb}''' - V_{O}^{\cdot\cdot})$ defect dipoles (where $M^{2+}$ represents the acceptor-doped ion in the Nb- site) in one surface layer during the pre-poling process, resulting in an asymmetrical distribution of defect dipoles in the two surface layers. The synergistic interaction between ferroelectric domains and defect dipoles in the surface layers induces this unidirectional electrobending, as evidenced by a butterfly-like symmetrical bipolar *S-E* curve with a giant apparent strain of 3.2%. These findings provide new insights into defect engineering strategies for developing advanced piezoelectric materials with large electroinduced displacements.

**Keywords:** electrobending, unidirectional bending, defect dipole, acceptor-doped, piezoceramic

**Introduction**

Piezoelectric materials, capable of converting electrical energy into mechanical strain, are widely employed as actuators in various fields, ranging from aerospace micro-thrusters to consumer electronics[1-3]. Since 2022, there has been a notable increase in reports of ultrahigh electrostrain (>1%)[4-8]. For instance, $0.94Bi_{0.5}Na_{0.5}TiO_3$–$0.06BaAlO_{2.5}$ ceramics exhibited an impressive electrostrain value of 1.12% at room temperature by incorporating oxygen vacancies[9]. In subsequent studies, this value was further enhanced to 1.6% in Sr/Nb-doped $Bi_{0.5}(Na_{0.82}K_{0.18})_{0.5}TiO_3$ lead-free textured piezoceramics through defect dipole engineering[10]. Additionally, Sr-modified $(K,Na)NbO_3$ ceramics achieved an electrostrain of 1.05% at room temperature[11], and later, $(K_{0.48}Na_{0.52})_{0.99}NbO_{2.995}$ ceramics exhibited an increased electrostrain of 2.1%[12], both improvements attributed to the presence of defect dipoles. Nevertheless, due to the inherent limitations of ion displacement within the crystal lattice, achieving electrostrain values exceeding 1% solely through lattice strain mechanisms—such as engineered domain configurations[13,14], electric-field-induced phase transitions[15,16], chemical disorder[17,18], and reversible switching of non-180° domains[19,20]—remains highly challenging.

Ranjan et al. subsequently demonstrated a significant correlation between ceramic thickness and measured strain, revealing a marked increase in apparent strain as the sample thickness decreases[21]. Notably, when the thickness is below 300 μm, the measured electrostrain can exceed 2%, a finding that has garnered widespread interest. Recently, several studies have reported a bending deformation phenomenon under an applied electric field in thin piezoceramics, which contributes to these ultrahigh apparent electrostrains[22-25]. In our previous work, we proposed a mechanism to explain this electric field-induced bending (electrobending) phenomenon, suggesting that inward-oriented defect dipoles in the surface layers generate a stress difference between the upper and lower surfaces under an external electric field, thereby inducing convex bending deformation along the negative electric field direction and resulting in asymmetrical bipolar strain-electric field (*S-E*) curves[24]. However, our recent studies have indicated that electrobending does not solely exhibit this bending mode.

In this study, a novel unidirectional bending deformation model, characterized by a symmetrical butterfly-like bipolar *S-E* curve, was observed in $0.95K_{0.5}Na_{0.5}NbO_3$-$0.05CaZrO_3$-

0.01CuO (KNNCZ-Cu) piezoceramic samples, which exhibited a giant apparent electrostrain of 3.2%. To further investigate the deformation behavior of these samples, strain measurement techniques were adjusted, yielding distinct *S-E* curves. This phenomenon is closely related to the $(M_{Nb}''' - V_O^{\cdot\cdot})$ defect dipoles (where $M^{2+}$ represents the acceptor-doped ion in Nb-site), part of which can be reoriented during the pre-poling process. After pre-poling, the distribution of defect dipoles in the upper and lower surface layers changes and becomes asymmetrical. The synergistic interaction between domains and defect dipoles in the surface layers results in the unidirectional electrobending, resulting in the symmetrical bipolar *S-E* curve. This discovery provides important insights into defect engineering and electrobending deformation, offering new possibilities for the design and application of piezoelectric materials with large electroinduced displacements.

**Results and discussion**

KNNCZ-Cu ceramics exhibit symmetrical *S-E* curves with significant apparent strain. To investigate the electrobending behavior, *S-E* curves for poled KNNCZ-Cu ceramic samples with varying thicknesses are shown in Fig. 1. The apparent electrostrain increases as the sample thickness decreases, ranging from 0.15% at 600 μm to 1.3% at 230 μm under an electric field of 50 kV/cm in Fig. 1a, indicating that bending deformation contributes to the increase in apparent strain. Additionally, the unipolar strain values are nearly identical under both positive and negative electric fields for samples with the same thickness, consistent with the symmetrical nature of the bipolar *S-E* curves, as shown in Figs. 1b and 1c.

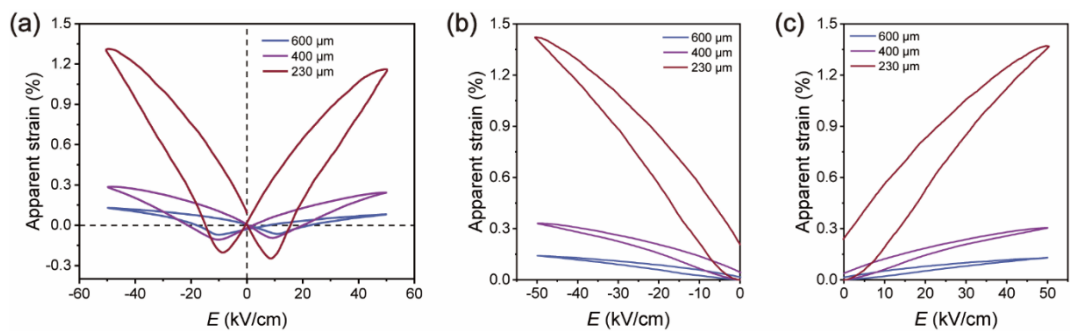

**Fig. 1 *S-E* curves of KNNCZ-Cu ceramics.** (**a**) Bipolar and (**b** and **c**) unipolar *S-E* curves of KNNCZ-Cu ceramics with different thicknesses.

The bipolar *S-E* curves of KNNCZ-Cu ceramics are influenced not only by sample thickness but also by the configuration of the measurement setup and the placement direction of the poled sample[22,23]. As shown in Figs. 2a-f, varying the test conditions allows modulation of the bipolar *S-E* curves and the apparent strain values. In test mode 1, when the pre-poling direction is oriented upwards, a symmetrical butterfly-shaped *S-E* loop with an apparent strain of 1% is observed (Fig. 2d). In test mode 2, where a copper shim is placed between the poled sample and the bottom holder, a significantly enhanced symmetric *S-E* loop is observed, with the maximum apparent strain increased to 2.8% (Fig. 2e), nearly three times the value obtained in test mode 1. Similarly, in test mode 3, where a copper shim is placed between the poled sample and the top holder, one enhanced symmetrical *S-E* curve is also observed, with the maximum apparent strain increased to 3.2% (Fig. 2f).

Based on the above results, we propose a unidirectional bending deformation model to explain the unique butterfly-shaped bipolar *S-E* curve. Specifically, regardless of the direction of the applied electric field, the thin ceramic sample bends convexly along the pre-poling direction. The ceramic disc bends convexly along the pre-poling direction and does not fully return to a flat state even after the pre-poling electric field is removed. Consequently, in test mode 1 (Fig. 2g), a convex pre-bending already exists before measurement (state $D_1$). When an electric field is applied, domain switching briefly alleviates the convex deformation (state $D_2$), resulting in a small negative strain, followed by further convex bending as the electric field increases (state $D_3$), leading to a large positive strain. Upon removal of the electric field, the ceramic gradually returns to its initial state (state $D_4$). In test mode 2 (Fig. 2h), the copper shim increases the contact area of the bottom holder with samples and can lift the top holder more effectively, thus leading to a higher apparent strain value (states $E_1$-$E_4$) compared to test mode 1. Similarly, as shown in Fig. 2i, when a downward pre-poling direction is combined with a copper shim placed between the top holder and the sample, an enhanced butterfly-shaped *S-E* curve can also be achieved, further confirming the unidirectional bending deformation in KNNCZ-Cu ceramics.

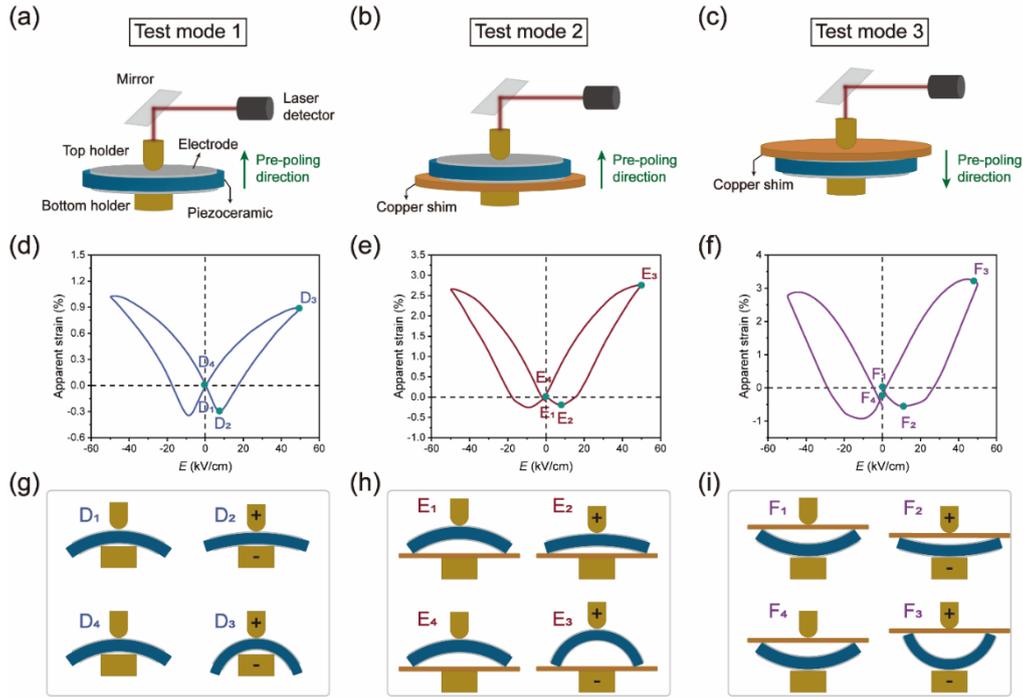

**Fig.2 Test modes with corresponding *S-E* curves and deformation processes.** (**a-c**) Schematic diagrams of three measurement modes. (**d-f**) Bipolar *S-E* curves of KNNCZ-Cu ceramics for the corresponding test modes. (**g-i**) Deformation models for test mode 1, 2, and 3, respectively.

Why do some acceptor-doped piezoceramics with a small thickness present this special bending deformation under an external electric field? A working mechanism is proposed to elucidate the phenomenon. During sample preparation, the acceptor-doped $Cu^{2+}$ ion and oxygen vacancies $V_O^{\cdot\cdot}$ can form <001>-oriented ($Cu_{Nb}''' - V_O^{\cdot\cdot}$) defect dipoles. Compared to the ceramic interior, the surface of ceramics exhibits a higher concentration of oxygen vacancies and, consequently, a higher concentration of defect dipoles in the surface layers. As discussed in our previous work, the defect dipoles in the two surface layers are more inclined to be oriented inward in order to reduce the energy[24], as shown in Fig. 3a. After a pre-poling process, the spontaneous polarizations ($P_S$) reorient along the pre-poling direction. Simultaneously, some defect dipoles in the upper surface layer, initially oriented opposite to the pre-poling field, are reoriented along the pre-poling direction under the applied field. Consequently, there are defect dipoles in the upper and lower surface layers that align with the pre-poling direction, though with a difference in their number. Upon removal of the pre-poling field, the ceramic sample exhibits a small convex bending along the pre-poling direction and cannot fully return to a flat state.

The synergistic interaction between defect dipoles and ferroelectric domains is responsible for the distinctive *S-E* curve and the unidirectional bending deformation. As depicted in Figs. 3b and 3c, the influence of defect dipoles and domains at various states ($O_1$, $O_2$, and A-D) in the *S-E* curve measured in test mode 1 is demonstrated. It should be noted that due to the difficulty in the reorientation of defect dipoles, which involves the vacancy migration and an energy-consuming diffusion process, the defect dipoles will retain their orientations during the measuring process[26,27].

From state $O_1$ to state A, under a relatively low positive electric field, the domains are switched from the upward to the horizontal direction, leading to a reduction in sample thickness and an extension in the in-plane direction, thereby alleviating the convex bending. As the electric field increases further to state B, the domains are nearly fully switched to the downward direction, resulting in an increase in thickness and a contraction in the in-plane direction. Meanwhile, since the defect dipoles are oriented opposite to the applied electric field, they are compressed under the field, causing the two surface layers to shorten in the thickness direction while extending in the in-plane direction. Due to the initial convex bending, the ceramic bends further in a convex manner, as shown in state B. Upon removal of the electric field, the convex bending is partially relieved, but some residual pre-bending remains (state $O_2$).

In state C, similar to state A, the domains are switched from the downward to the horizontal direction, resulting in a decrease in sample thickness and alleviating the convex bending. As the electric field further increases to state D, the domains are nearly fully switched to the upward direction and are extended. Since the orientation of the defect dipoles aligns with the applied electric field, they are stretched under the field, causing an increase in thickness and a contraction in the in-plane direction. However, due to the difference in the number of defect dipoles between the upper and lower surface layers, the contraction in the lower surface layer in the in-plane direction is greater than that in the upper one, leading to further convex bending. Upon removal of the electric field, the convex bending is partially relieved, but some residual pre-bending remains (state $O_1$).

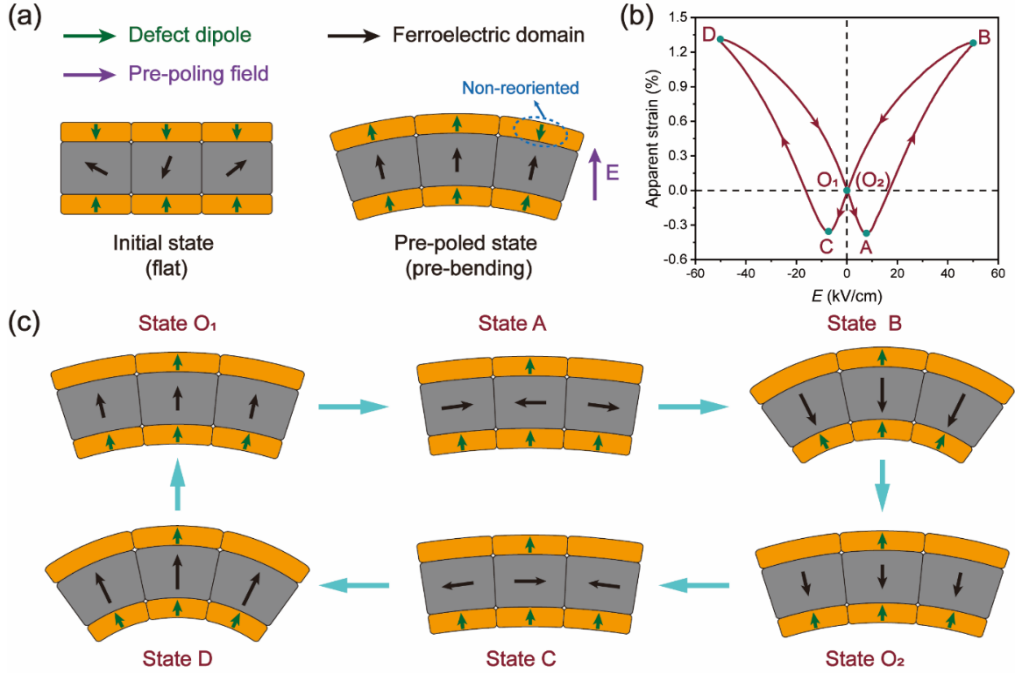

**Fig. 3 Mechanism for the unidirectional electrobending.** (a) Orientation of surface layer defect dipoles and ferroelectric domains in the initial state and pre-poled state. (b) Bipolar *S-E* curve of KNNCZ-Cu ceramics in test mode 1. (c) Different states represented in the *S-E* curve.

To better understand the behavior differences between <001>-oriented $(Cu'''_{Nb} - V_O^{\cdot\cdot})$ and <110>-oriented $(V'_A - V_O^{\cdot\cdot})$ defect dipoles in surface layers, we present a schematic illustration of 180° reorientation for these two types of defect dipoles (Fig. 4a). For the $(Cu'''_{Nb} - V_O^{\cdot\cdot})$ defect dipole, oxygen vacancy migration requires two steps within a single unit cell to achieve a 180° reorientation under a reverse electric field. In contrast, the $(V'_A - V_O^{\cdot\cdot})$ defect dipole can also undergo 180° reorientation in two steps, but this requires crossing two unit cells, resulting in a longer migration distance. The reorientation energy barriers of these two defect dipoles were calculated using a 3 × 3 × 3 supercell for density functional theory (DFT) calculations. As illustrated in Fig. 4b, reorienting the $(Cu'''_{Nb} - V_O^{\cdot\cdot})$ defect dipole requires approximately 0.89 eV, while the $(V'_A - V_O^{\cdot\cdot})$ defect dipole requires 1.56 eV. Furthermore, the electrostatic energy between $(V'_A - V_O^{\cdot\cdot})$ defect dipole and spontaneous polarization is approximately three times greater than that between $(Cu'''_{Nb} - V_O^{\cdot\cdot})$ defect dipole and spontaneous polarization in KNN ceramics[12], indicating higher energy consumption and enhanced stability compared to the $(Cu'''_{Nb} - V_O^{\cdot\cdot})$ defect dipole. Consequently, the $(Cu'''_{Nb} - V_O^{\cdot\cdot})$ defect dipoles are more readily reoriented during the pre-poling process, while the $(V'_A - V^{\cdot\cdot}_O)$ defect dipoles tend to remain stable. These differences in reorientation energy barriers among defect dipoles lead to the different bending deformation behaviors and *S-E* curve characteristics observed in thin piezoelectric materials under applied electric fields.

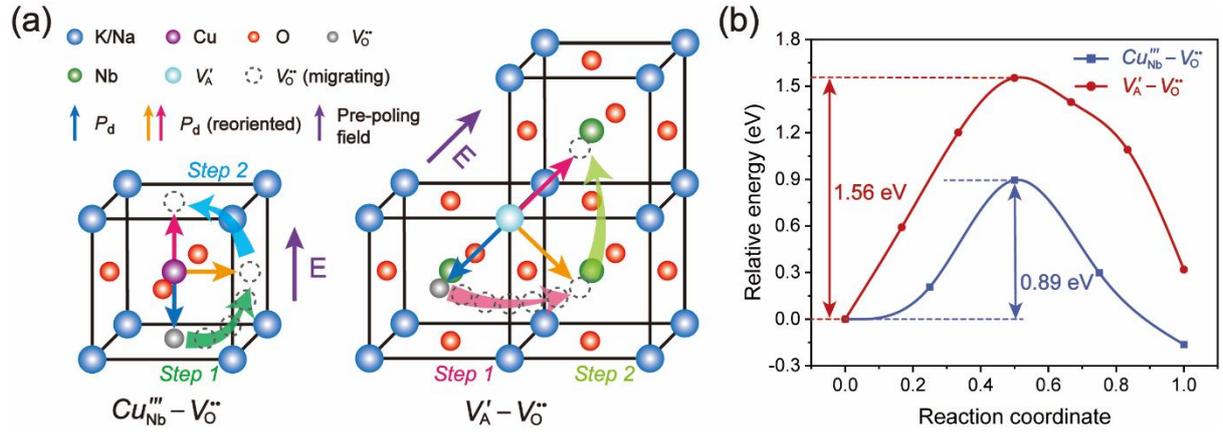

**Fig. 4 First-principles calculations.** (**a**) Schematic diagrams of 180° reorientation and (**b**) reorientation energy barriers of $(Cu'''_{Nb} - V^{\cdot\cdot}_O)$ and $(V'_A - V^{\cdot\cdot}_O)$ defect dipoles, respectively.

## Conclusions

In summary, we present a novel unidirectional electrobending behavior in acceptor-doped $K_{0.5}Na_{0.5}NbO_3$ ceramics, wherein convex bending consistently occurs along the pre-poling direction, regardless of the applied electric field direction. This behavior is reflected in a symmetrical butterfly-like symmetrical bipolar *S-E* curve, demonstrating an impressive apparent strain of 3.2%. The unique characteristics of this unidirectional electrobending are attributed to the asymmetrical distribution of defect dipoles in the surface layers following a pre-poling process, combined with domain switching. These insights not only advance our understanding of defect engineering but also offer significant potential for developing advanced materials with large electroinduced displacements.

## Methods

Sample preparation

The $0.95K_{0.5}Na_{0.5}NbO_3$-$0.05CaZrO_3$-$0.01CuO$ ceramics were prepared by a conventional high temperature solid state sintering method. Metal oxides and carbonate powders of $K_2CO_3$ (99.5%), $Na_2CO_3$ (99.8%), $Nb_2O_5$ (99.99%), $CaCO_3$ (99.0%), $ZrO_2$ (99.99%) and CuO (99.9%) were used as raw materials. After being weighted according to a stoichiometric ratio, they were mixed and ball-milled for 12 h in a nylon jar with zirconia balls as the medium. The dried powders were calcined at 850 °C for 3 h, followed by a ball mill for another 12 h. The granulated powder was uniaxially pressed into discs of 10 mm in diameter and 1 mm in thickness under 200 MPa using polyvinyl butyral (PVB) as a binder. After burning off the organic substances, the green pellets were sintered at 1080°C for 4 h in covered alumina crucibles. Then the sample pellets were mechanically thinning to below 0.3 mm. Subsequently, ceramics were printed with silver ink on both sides and fired at 720 °C for 15 min to form electrodes.

Electrical measurements

Before electrical measurement, the samples were pre-poled in silicone oil with a direct-current (DC) electric field of 40 kV/cm for 30 minutes at room temperature. The *P-E* loops and *S-E* curves were measured by a piezoelectric evaluation system (aix-ACCTF Analyzer 2000E, Aachen, Germany).

First-principles canulations

First-principles calculations were performed by the density functional theory (DFT) using the projector augmented plane-wave method[28] within the Vienna Ab initio Simulation Package (VASP)[29,30]. The generalized gradient approximation (GGA) was used in the scheme of Perdew-Burke-Ernzerhof (PBE) to describe the exchange-correlation functional[31]. The diffusion energy barriers at different adsorption sites were estimated by using the Nudge Elastic Band (CI-NEB) method[32]. The cut-off energy for the plane wave was set to 480 eV. The energy criterion was set to $10^{-4}$ eV in the iterative solution of the Kohn-Sham equation. All the structures were relaxed until the residual forces on the atoms had declined to less than 0.05 eV $Å^{-1}$. To prevent interaction between periodic units in the vertical direction, a vacuum space of 20 Å was employed.

**Acknowledgments:** This work was supported by the National Key R&D Program of China (grant no. 2021YFA0716500), the National Natural Science Foundation of China (52172135), the Youth Top Talent Project of the National Special Support Program (2021-527-07), the Leading Talent Project of the National Special Support Program (2022WRLJ003), and the Guangdong Basic and Applied Basic Research Foundation for Distinguished Young Scholars (grant nos. 2022B1515020070 and 2021B1515020083).

**Data and materials availability:** All data are available in the main text.